# Direct Observation of Polarization vs. Thickness Relation in Ultra-Thin Ferroelectric Films

Rene Meyer, Arturas Vailionis, and Paul. C. McIntyre

Geballe Laboratory for Advanced Materials and Materials Science and Engineering Department, Stanford University, Stanford, CA

**Abstract**
A reduction of polarization in ultra-thin ferroelectric films appears to be fundamental to ferroelectricity at the nanoscale. For the model system $PbTiO_3$ on $SrTiO_3$, we report observation of the polarization vs. thickness relation. Distinct periodicity changes of ferroelectric domains obtained from x-ray diffraction and total energy calculations reveal a linear lowering of the polarization below a critical thickness of ~12 nm. Independent polarization and tetragonality measurements provide insight into the fundamental relation between polarization and tetragonality in nanoscale ferroelectrics.

Thin ferroelectric films of thickness less than 10 nm have recently been the focus of both experimental and theoretical efforts to address fundamental questions about ferroelectricity on the nanoscale. Due to the collective nature of ferroelectric phenomena, ferroelectric material properties are expected to change below a critical film thickness or probe volume. Various materials such as oxides and polymers have been studied in order to explore the transition from ferroelectric to a super-paraelectric state and to obtain the critical size at which ferroelectricity may be suppressed [1-5].

Measuring the switchable polarization by electrical characterization techniques provides the most direct and quantitative results on ferroelectricity. On the nanoscale, however, very large leakage currents present in many materials systems requires the adoption of alternative methods to measure polarization. Electrical studies of epitaxial $BaTiO_3$ films show ferroelectricity down to 3 nm [6]. Similar investigations of the model material $PbTiO_3$ are not possible due to high leakage currents, which compromise the measurement of ferroelectric response and displacement current. Instead, changes in the unit cell tetragonality with film thickness are used as an *indirect* measure of the spontaneous polarization. A sharp reduction of tetragonality for thicknesses below 5 nm was observed in thin PTO layers grown on conductive $SrTiO_3$(001) substrates [7]. Lichtensteiger *et al.* related the decrease of tetragonality to a reduction of spontaneous polarization through first-principles-based model Hamiltonian approach. They concluded that a lowering of tetragonality implies the existence of a critical thickness below which ferrolelectricity vanishes [7].

Recent *in situ* synchrotron x-ray scattering studies utilized the existence of ferroelectric domains, also referred to as stripe domains, as *direct* evidence of ferroelectricity [8]. Stripe domains produce a distinct x-ray diffraction intensity pattern around PTO Bragg reflections. This work has demonstrated that, at elevated temperatures, PTO layers are ferroelectric down to 3 unit cells [8,9]. Large ionic displacement together with the availability of high quality (001) oriented epitaxial films make this material a very suitable model system for studying ferroelectricity by x-ray diffraction (XRD). Temperature dependent observations of stripe domains in thin PTO films using synchrotron x-ray scattering by Fong *et al.* [9] suggest the existence of three different domain configurations: two multi-domain states and one mono-domain state. While it has been suggested that a variation in the domain density at elevated temperatures originates from a change from an unscreened to a screened ferroelectric surface (due to charge transfer from surface adsorbates), it remains unclear, why ferroelectric domains should not be observed at room temperature using synchrotron radiation.

In this letter, we demonstrate that x-ray diffraction techniques can be utilized to study the polarization as a function of film thickness *quantitatively* in relatively conductive materials, in which ferroelectric properties cannot be characterized electrically. By combining measurement with theory we have, for the first time, extracted polarization vs. thickness and polarization vs. tetragonality relations in nanoscale PTO single crystal thin films. The observed linear decay of the spontaneous polarization appears to be fundamental to ferroelectricity in this nanoscale system. Characterization of ferroelectric domains at room temperature by a laboratory x-ray source and a comparative study using synchrotron light indicate that the disappearance of domains in ultra-thin PTO films at low temperatures reported by Fong *et al.* may be caused by prolonged exposure to synchrotron irradiation, which may have altered the domain configuration in this temperature range.

Our experiments exploit the fact that ferroic materials form domains of particular sizes in order to minimize the free energy of the system. Under equilibrium conditions, the domain periodicity gives an insight into the relationships among electrostatic, elastic and domain wall energies as a function of film thickness, polarization, strain, temperature and the particular screening conditions of the polarization charge at the surface and at the film/substrate interface. If the system boundary conditions are well known, determination of domain period and film thickness allows us to back-calculate the polarization at a given thickness.

Domains in thin ferroelectric PTO layers are present as $180°$ stripe domains due to out-of-plane displacement of the Pb/Ti ions in the tetragonal unit cell. To study domain formation as a function of film thickness, $PbTiO_3$ films were grown in a cold wall MOCVD chamber using tetraethyl lead $Pb(C_2H_5)_4$ and titanium isopropoxide $Ti(OC_3H_7)_4$ metal precursors on insulating $SrTiO_3(001)$ substrates. At a deposition temperature of 600°C and a molecular oxygen pressure of 1.5 Torr, films grew with a rate of 2.5 Å/min. Due to the fact that cooling under oxygen deficient conditions can significantly affect the c-axis parameter, films were cooled in a defined oxygen atmosphere [10]. X-ray diffraction experiments were employed to characterize ferroelectric domain periodicity, structural quality and PTO film thickness using a PANalytical X'Pert materials research diffractometer at the Stanford Nanocharacterization Laboratory as well as beam line 7-2 at the Stanford Synchrotron Radiation Laboratory. The stripe domain periodicity is determined from the lateral separation of the first-order satellites around PTO(203) and (403) Bragg reflections observable at room temperature. Layer thicknesses for all films were estimated from x-ray reflectivity scans. Fitting the observed reflectivity data to the calculated profiles revealed that the PTO layers exhibit atomically abrupt interfaces and a surface roughness less than 3 Å. Reciprocal space maps confirm that PTO layers were grown in a fully coherent fashion to the underlying $SrTiO_3(001)$ substrate. The c-axis lattice parameter was determined from the position of the PTO(004) reflection. Tetragonality was calculated from c-axis and a-axis lattice parameters. For layers with thicknesses below 4 nm, the c-axis lattice parameter was obtained from asymmetrical (403) scans using a synchrotron radiation source.

The stripe domain periodicity $\Lambda$ is a direct measure of the spontaneous polarization in ferroelectric materials. For a constant polarization, an expected $\Lambda \propto d^2$ law is observed, where $d$ is the film thickness [8]. A deviation from this law indicates a change in the polarization with film thickness. While the electrostatic energy as a function of the polarization scales with a power 2 law, the domain wall energy $\gamma$ scales to the power 3 [11]:

$$\gamma(P_S) = \gamma^0 \left(\frac{P_S}{P_S^0}\right)^3. \tag{1}$$

If the material's polarization increases, the domain wall energy will increase more than the electrostatic energy. Hence, domains will grow as polarization increases for a given film thickness. We apply a two-dimensional (2D) numerical model to quantify this relation and to extract the polarization by calculating the total energy of the system as the sum of electrostatic energy $F_{es}$ and domain wall energy $\gamma$. Figure 1 illustrates the model system of ferroelectric and dielectric layers. An additional vacuum dielectric layer of variable thickness $d_{vacuum}$ between the ferroelectric $d_{FE}$ and the top electrode allows one to calculate the domain periodicity for different screening efficiencies of the polarization charge at the ferroelectric surface. At large thicknesses of the vacuum layer, we obtain the domain periodicity for an unscreened surface. For the calculation presented here, it is assumed that the free surface is fully screened at room temperature as reported earlier by Fong *et al.* [9] and $d_{vacuum}$ is set to zero. The thickness of the substrate layer is chosen to be large compared to the ferroelectric layer, so that the electrostatic energy is not a function of $d_S$.

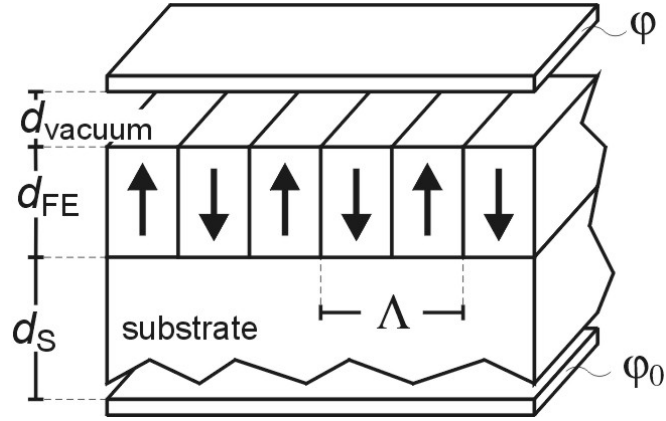

Fig. 1: Model structure of ferroelectric and dielectric layers.

Minimizing the total energy as a function of spontaneous polarization $P_S$ allows one to estimate the domain width $\Lambda$ vs. polarization relation for different thicknesses of the ferroelectric $d_{FE}$. Possible contributions from the elastic energy have not been taken into consideration in this calculation [11]. The electrostatic energy $F_{es}$ is given by

$$F_{es} = \iiint \frac{1}{2}\rho\varphi dV + \iint \frac{1}{2}\sigma\varphi d\Sigma , \qquad (2)$$

where $\rho$ denotes the space charge density in the volume element $dV$, $\sigma$ is the surface charge density of the surface element $d\Sigma$ and $\varphi$ denotes the electrostatic potential. The potential distribution $\varphi$ inside the ferroelectric layer and the substrate is calculated by solving Poisson's equation for a heterogeneous anisotropic medium:

$$\nabla\varepsilon_r\nabla\varphi + \varepsilon_r\nabla^2\varphi = -\frac{\rho}{\varepsilon_0} \qquad (3)$$

In Eq. 3, $\varepsilon_0$ denotes the dielectric constant of the free space and $\rho$ is the local charge density of the free and the bound charges only from the spontaneous polarization. It is assumed that for thick films, the polarization saturates at $P_{S,max} = 85\mu C/cm^2$ [12]. This polarization value is consistent with the results of ferroelectric switching current measurements performed on our thicker PTO films with conductive STO substrates and Pt top electrodes. In the ferroelectric, $\varepsilon_r$ refers to the in-plane and out-of-plane component of the dielectric constant $\varepsilon_\perp$ =1200 and $\varepsilon_\parallel$ =150. For the substrate, we assume a homogeneous dielectric constant $\varepsilon_s$ =300. The field dependence of the dielectric constants is not considered here to simplify the problem. To solve the differential equation, we apply periodic in-plane boundary conditions and Dirichlet type out-of-plane boundary conditions for the potential.

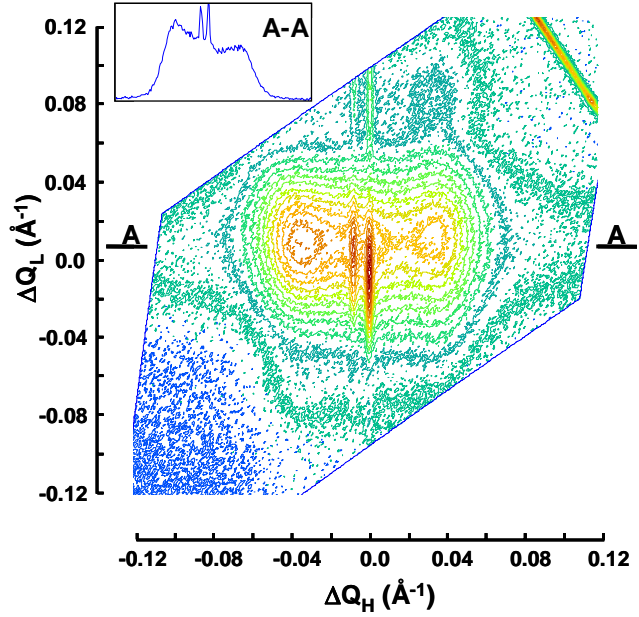

Fig. 2: (H0L) reciprocal space map around the (203) Bragg reflection of a 12 nm PTO layer measured with X'Pert materials research diffractometer at room temperature.

Figure 2 shows the (H0L) reciprocal space map around the (203) Bragg reflection of a 12 nm PTO layer measured with X'Pert materials research diffractometer at room temperature. The main (203) Bragg peak consists of the K$\alpha_1$ and the K$\alpha_2$ reflections. Layer peak and thickness fringes both show first order satellites at $\Lambda = 2\pi/\Delta Q$ from the Bragg peak. The horizontal alignment of the satellites around the main diffraction peak is consistent with a 180° in-plane stripe domain structure in the PTO film. Figure 3 displays the stripe domain periodicity $\Lambda$ as a function of the film thickness $d_{Fe}$ on a double logarithmic scale. Two distinct $\Lambda(d_{Fe})$ relations are observed: for films thicker than ~12 nm, we find a slope of ½, as expected, and this does not vary with film thickness [8]. Films of 12 nm thickness and less, however, show a slope slightly greater than ½, making evident that the polarization is not constant in this thickness range [13]. Reduced polarization values make it energetically more favorable to increase the domain wall density. We conclude that the spontaneous polarization changes as a function of film thickness and decays as films get thinner. For films thicker than ~12 nm, the polarization saturates. A gradual reduction of the stripe domain periodicity indicates no abrupt disappearance of ferroelectricity down to 4 unit cells. Data obtained from the synchrotron source agree well with the results of the laboratory source experiment for a 3.5 nm thick film, and show a consistent domain periodicity vs. thickness trend for a 1.6 nm thick PTO layer. Preliminary comparative observations of domain structure measured by laboratory and synchrotron sources reveal a significant effect of prolonged exposure of the PTO films to synchrotron light on the stripe domain periodicity. We found indications that domains grow, during PTO film exposure to synchrotron radiation for several hours. We conclude that the mono-domain state of the PTO films at room temperature reported by Fong *et al.* may have been caused by synchrotron radiation exposure.

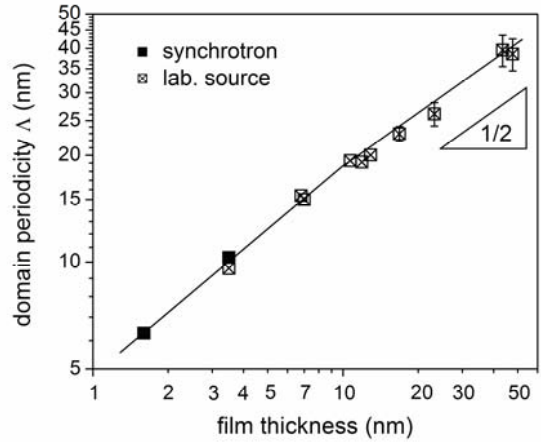

Fig. 3: Stripe domain periodicity as a function of the film thickness from laboratory x-ray source (crossed squares) and synchrotron source (filled squares).

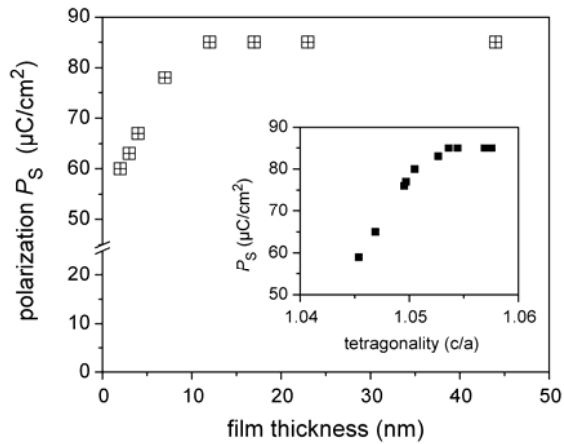

Fig. 4: Spontaneous polarization back-calculated from domain periodicity. Inset: Back-calculated polarization versus film tetragonality.

We apply our numerical model to quantify the polarization loss as a function of film thickness by back-calculating the polarization for each particular film thickness from the experimental data for stripe domain periods. Spontaneous polarization as a function of the film thickness estimated from the domain periodicity is displayed in Fig. 4. We find a constant polarization for films thicker than ~12 nm and a linear decay of the polarization for film thicknesses of ~12 nm down to 4 unit cells. The reduction of the polarization is less pronounced than reported in earlier works [7] and the 4 unit cell thick film still has a high polarization of 60 μC/cm². The result is surprising, and indicates that there may be no intrinsic thickness limit, below which ferroelectricity is suppressed.

To substantiate the thickness, at which the material polarization becomes a function of film thickness, studies of the PTO tetragonality were performed. The complementary measurements of polarization and tetragonality provide insight into the relationship between unit cell geometry and ferroelectric polarization on the nanoscale. The onset of tetragonality reduction with film thickness correlates reasonably well with the change of the calculated polarization. The c-axis/a-axis ratio dependence on thickness is illustrated in Fig. 5.

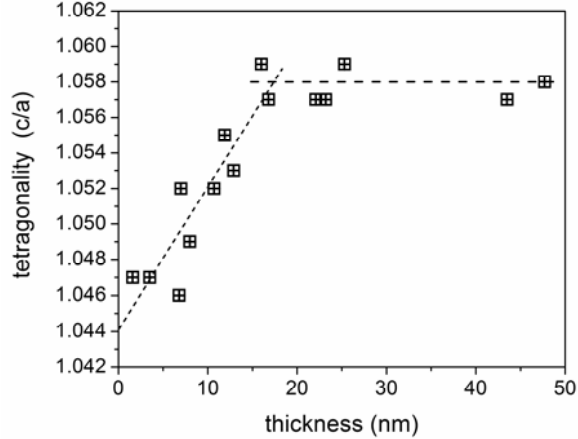

fig. 5: Tetragonality (c-axis/a-axis lattice parameter ratio) determined from PTO (004) and (403) Bragg peak position as a function of the film thickness.

As the PTO layer thickness is reduced from ~16 nm to 1.6 nm, the tetragonality changes from ~1.057 to ~1.047. The reduction occurs gradually and no indications are found for an abrupt decrease in tetragonality that might point to a suppression of ferroelectricity below a certain critical thickness. Films thicker than ~16 nm have a constant c-axis lattice parameter of 4.13 Å. A reduction of the polarization with reduced tetragonality can be explained qualitatively by the reduced distance between the potential minima of the double potential in the center of the unit cell, which will cause smaller ionic displacements. Generic linear fits to the thickness-independent and thickness-dependent regimes of the tetragonality vs. PTO film thickness data allow us now to quantify the fundamental relation between polarization and tetragonality under constrained geometries. The obtained linear relation for films below 12 nm thickness and the saturation for films thicker 12 nm are displayed as an inset to Fig. 4. Results show that the reduction of the polarization is small and does not indicate a clear ferroelectric size limit. The ferroelectric limit reported for PTO in earlier studies for film thicknesses of 4 unit cells and larger might therefore be of extrinsic, rather than intrinsic nature. In future studies, we plan to apply the presented method to films with different defect concentrations to study the effects of point defects on polarization at a given tetragonality.

In summary, we report the fundamental relation between polarization and film thickness in ultra-thin PTO films grown epitaxially on SrTiO$_3$(001) substrates. The direct determination of room temperature polarization is achieved by combining x-ray diffraction measurements of 180° stripe domains with theory. A deviation in the film thickness vs. domain periodicity relation from the classical scaling law provides the first evidence that the polarization decreases with decreasing thickness below 12 nm. Total energy calculations are utilized to quantify this effect and to estimate the spontaneous polarization as a function of film thickness. A linear reduction of the polarization is found for films of 12 nm thickness and less. No abrupt ferroelectric/dielectric transition is observed for layers down to 4 unit cells. The observed reduction of the polarization coincides with a reduction of the c-axis lattice parameter for layer thicknesses less than 16 nm. Independent estimation of the polarization-thickness relation from stripe domain periodicities and the tetragonality-thickness relation by x-ray analysis allows, for the first time, quantification of the tetragonality-polarization relationship in nanoscale ferroelectrics.

Portions of this research were carried out at the Stanford Synchrotron Radiation Laboratory, a national user facility operated by Stanford University on behalf of the U.S. Department of Energy, Office of Basic Energy Sciences. R. Meyer and P.C. McIntyre acknowledge support from the National Science Foundation (DMR grant 0205949), Toshiba Corporation, and the Stanford Non-Volatile Memory Technology Initiative.